
\magnification=\magstep1
\openup 1\jot
\def\bo{ { \sqcup\llap{ $\sqcap$} } }
\overfullrule=0pt       

\font\cat=cmr7

\def\real{I\negthinspace R}
\def\zed{Z\hskip -3mm Z }

     
\newcount\refno
\refno=0
\def\nref#1\par{\advance\refno by1\item{[\the\refno]~}#1}
     
\def\book#1[[#2]]{{\it#1\/} (#2).}
     
\def\apj#1 #2 #3.{{\it Ap.\ J.\ \bf#1} #2 (#3).}
\def\cqg#1 #2 #3.{{\it Class.\ Quant.\ Grav.\ \bf#1} #2 (#3).}
\def\npb#1 #2 #3.{{\it Nucl.\ Phys.\ \rm B\bf#1} #2 (#3).}
\def\phrep#1 #2 #3.{{\it Phys.\ Rep.\ \bf#1} #2 (#3).}
\def\plb#1 #2 #3.{{\it Phys.\ Lett.\ \bf#1\/}B #2 (#3).}
\def\pr#1 #2 #3.{{\it Phys.\ Rev.\ \bf#1} #2 (#3).}
\def\prsa#1 #2 #3.{{\it Proc.\ Roy.\ Soc.\ \rm A\bf#1} #2 (#3).}
\def\prd#1 #2 #3.{{\it Phys.\ Rev.\ \rm D\bf#1} #2 (#3).}
\def\prl#1 #2 #3.{{\it Phys.\ Rev.\ Lett.\ \bf#1} #2 (#3).}
\def\rprog#1 #2 #3.{{\it Rep.\ Prog.\ Phys.\ \bf#1} #2 (#3).}
     
\hbox{ }
\rightline {DTP/95/61}
\rightline {hep-th/9510202}
\rightline {October 1995}
\vskip 1truecm

\centerline{\bf COSMIC p-BRANES}
\vskip 1truecm

\centerline{Ruth Gregory\footnote{$^\spadesuit$}{\sl Email:
dma0rag@gauss.dur.ac.uk}}
\vskip 2mm
\centerline{ \it Centre for Particle Theory, }
\centerline{\it University of Durham, Durham, DH1 3LE, U.K.}

\vskip 4mm
\centerline{\cat ABSTRACT}
\vskip 4mm

{
\openup -1 \jot

We consider the metrics for cosmic strings and $p$-branes in
spacetime dimension $N>4$, that is, we look for solutions to 
Einstein-Maxwell-Dilaton gravity
in $N$-dimensions with boost symmetry in the $p$-directions along the brane.
Focussing first in detail on the five dimensional 
uncharged cosmic string we discuss the
solution, which turns out to have a naked singularity on the brane, as well as
considering its Kaluza-Klein reduction. We show how singularities may be
avoided with particular core models. We then derive the general
uncharged $p$-brane solution in arbitrary dimension. Finally, we consider
an Einstein-Maxwell-Dilaton action, with arbitrary value of the dilaton
coupling parameter, deriving the solutions for electrically and magnetically
charged branes, as well as a class of self-dual branes. 

\openup 1\jot
}



\vfill\eject
\footline={\hss\tenrm\folio\hss}

\noindent{\bf 1. Introduction.}

\vskip 2mm

Although it is true that on the large scale we seem to live in a
four-dimensional world, the idea that spacetime might have more than four
dimensions, with those extra dimensions curled up on a very small scale, has
always had some attraction, most recently in the context of string theory.
From a gravitational point of view, one of the interesting features of having
more dimensions is that event horizons can have more interesting
topologies. Instead of being restricted to spherical, or hyperspherical,
surfaces, exact solutions exist [1,2] which have extended event horizons
such as $S^2\times$\real$^7$ etc.
In fact, any vacuum solution to Einstein's equations in N-dimensions
can be trivially extended to an (N+1)-dimensional solution simply by
adding in a flat direction. The solutions of Gibbons, Maeda, Horowitz and 
Strominger (GMHS) are less trivial in that they  represent charged extended
objects in Einstein-dilaton gravity, however one notable feature of these
solutions is their lack of boost symmetry. 
The uncharged `brane' has a composite metric of the aforementioned form: a 
D-dimensional black hole times a p-dimensional flat euclidean space.
Although the metric of the charged brane is more complex, the magnetically
charged branes with a specific dilaton coupling appropriate to low energy
string gravity do take such a form[2].
Curiously, for all brane types, boost symmetry is restored in the extremal
limit. Another interesting feature of the GMHS solutions is that most of the
exact solutions are unstable [3], excepting those that are extremal [4]. Since
the energy-momentum of the charge-field and dilaton does have boost
symmetry, it is tempting to suggest that these instabilities are
connected with the fact that the spacetime symmetries do not correspond
to the source symmetries. This leaves us with the question:
How does the metric of a boost symmetric brane
behave?

From a completely different viewpoint,
in cosmology we are often interested in the gravitational properties of cosmic
strings and other defects, since these may have relevance for structure
formation in the early universe (for recent reviews see [5] and [6]). 
If there are extra dimensions, it is
interesting to query how these might effect the metric of a cosmic string, say.
For example, suppose spacetime is topologically \real$^4\times S^1$, as in
Kaluza-Klein theory. If our string is infinite in the \real$^4$ part of the
spacetime, we might hope that the metric is essentially unaffected, and indeed
one can see that this is the case from considering the four-dimensional
Kaluza-Klein theory. However, what if the string winds around the internal
$S^1$? This would appear to be a point-like, presumably uncharged, source and
hence a Schwarzschild metric such as one might obtain from dimensionally
reducing the solutions in [1,2]. However, as we will show, such an argument
ignores the crucial feature of a cosmic string (and more general $p$-brane
defects formed as solitons in some field theory) 
which is the boost symmetry on the brane. The five-dimensional
(uncompactified) Schwarzschild $\times$\real~metric cannot be the metric of a
vortex in five dimensions precisely because it does not have boost symmetry.
What then does a five-dimensional string look like?

Finally, it has recently been argued that cosmic strings might be 
unstable to black hole pair creation along their length[7-10]. 
Although the topology of the vacuum manifold of a cosmic string (or other
defect) generally protects against decay; if one allows a {\it spacetime} 
topology changing process, such as black hole pair creation, 
then the usual stability arguments can be sidestepped.
Such a process appears to be completely democratic, if it can occur (see [10]
for a discussion of caveats) then it will occur, whether or not the defect
is of Bogomolnyi type\footnote{$^*$}{
We are using the term Bogomolnyi defect in the usual sense here, 
in that it refers to a specific field theory in which
the second order equations can be reduced to first order by virtue
of a special choice of coupling constants[11]. Such field theories are
generally supersymmetrizable.}.
The instability in four-dimensions relies on the C-metric[12], an exact
solution of the vacuum Einstein equations representing two black holes 
uniformly accelerating apart, connected to infinity by a pair of conical
deficit singularities. The conical singularity of the C-metric is then
replaced by the snub-nosed cone of the cosmic string[10]. Suppose a similar
decay process exists for higher dimensional strings and branes, then by the 
principle of democracy, both Bogomolnyi and non-Bogomolnyi defects would
be susceptible to decay. However, a necessary first step to any such 
investigation is an understanding of the higher dimensional analogue
of the conical singularity in four dimensions.

In this paper, we attempt to answer these questions. We look for metrics which
have the form
$$
\hbox{(D-Hole)}\times \hbox{p-boost directions}
$$
and might therefore correspond to $p$-branes in general dimensions. 
We consider uncharged and charged branes in $N$-dimensional  
gravity. We find the general form of the metric which turns out generically 
to have a null naked singularity at the core. We will argue that this is
a property only of the exact solution, and can be smoothed out by a core
model, rather like the four-dimensional cosmic string core smooths out the apex
of the conical singularity[13,14].

The layout of the paper is as follows. In the next section we focus on the
five-dimensional uncharged string. We derive the metric, discussing the
singularity and core models, finally considering the Kaluza-Klein reduction of
our solution. In section three we generalise our results to 
uncharged $p$-branes in
$N$-dimensions. In section four, we consider the charged five-dimensional
string, using an 
Einstein-Maxwell-Dilaton gravity with arbitrary dilaton coupling.
We consider the general case in section five, giving 
a few string-motivated examples. We then consider the electrically charged
solutions and a special class of self-dual solutions.
Finally we sum up our results.

\vskip 4mm

\noindent{\bf 2. The Five-Dimensional String.}

\vskip 2mm

Let us start by examining a string in five dimensions since this is the first
non-trivial scenario to consider, and it is useful for visualisation. We first
look for a vacuum solution, since any real cosmic string ought to asymptote
this form.

One can assume (or show, with reference to a specific core model) that the
5-dimensional boost symmetric metric will take the form
$$
ds^2 = A^2(dt^2-dz^2) - B^2 dr^2 - C^2(d\theta^2+\sin^2\theta d\phi^2),
\eqno (2.1)
$$
which has the vacuum Einstein equations
$$
\eqalignno{
R^0_0&= B^{-2} \left [ {A''\over A} - {A'B'\over AB} 
+ \left ({A'\over A} \right )^2 + 2 {A'C'\over AC} \right ] = 0 & (2.2a) \cr
R^r_r &= 2 B^{-2} \left [ {A''\over A} + {C''\over C} 
- {B'\over B}\left ( {A'\over A} + {C'\over C} \right )
\right ] = 0 & (2.2b) \cr
R^\theta_\theta &=  B^{-2} \left [ {C''\over C}
+ {C'\over C} \left ( {C'\over C}+ 2{A'\over A} - {B'\over B} \right ) \right ]
- {1\over C^2} = 0. & (2.2c) \cr
}
$$

Obviously there is still coordinate freedom in the metric (2.1), and we choose
to restrict it in part by setting
$$
B=A^{-2}.
\eqno (2.3)
$$
This somewhat unusual choice was motivated by trying to take account of the
extra ``$dz^2$'' piece multiplying $A^2$, but turns out to give the simplest
form for the solution. Obviously we need boundary conditions in order to
solve (2.2), and since we are finding an asymptotic solution we impose
boundary conditions at infinity, demanding that the spacetime be asymptotically
Minkowskian in the four-dimensional sections transverse to the string,
in other words
$$
C\sim r \;\;\; ; \;\;\; A\to 1 \;\;\; {\rm as} \;\; r\to \infty.
\eqno (2.4)
$$

With the substitution (2.3), the Einstein equations can be written in the form
$$
\eqalignno{
((A^4)'C^2)' &= 0 & (2.5a) \cr
\left ({A'\over A} \right )^2 + \left ({C'\over C} \right )^2 +
\left ({A'C'\over AC} \right ) &= {1\over C^2}& (2.5b) \cr
(A^4(C^2)')' &=2 .& (2.5c) \cr
}
$$
Hence by direct integration
$$
\eqalignno{
(A^4)'C^2 &= 4a_0 & (2.6a) \cr
A^4(C^2)' &= 2r+2c_0 & (2.6b) \cr
}
$$
and thence
$$
A^4C^2 = r^2 + 2(c_0+2a_0)r + b_0=(r-r_+)(r-r_-), \eqno(2.6c) 
$$
where
(2.5b) gives
$$
b_0 = c_0^2 + a_0^2 + 4a_0c_0\Rightarrow r_\pm = -c_0 -2a_0\pm\sqrt{3}a_0.
\eqno (2.7)
$$
Choosing the origin of our $r$-coordinate to set $r_-=0$
($c_0= -(2+\sqrt{3})a_0$), it is not difficult to show that the
metric of the string is given by
$$
ds^2 = \left ( 1 - {r_+\over r} \right )^{1\over \sqrt{3}}(dt^2-dz^2)
- \left ( 1 - {r_+\over r} \right )^{-2\over \sqrt{3}}dr^2
- r^2 \left ( 1 - {r_+\over r} \right) ^{1-{2\over \sqrt{3}}}
(d\theta^2+\sin^2\theta d\phi^2).
\eqno (2.8)
$$

What are the important features of this metric?
First of all, it is
asymptotically flat in the 4-dimensional sense, asymptoting
$$
ds^2 \simeq \left ( 1 - {r_+\over\sqrt{3}r} \right ) (dt^2-dz^2)
- \left ( 1 + {2r_+\over\sqrt{3}r} \right ) dr^2 
-r^2 \left ( 1 + {(2-\sqrt{3})r_+\over \sqrt{3}r} \right ) d\Omega^2_{I\!I}
\eqno (2.9)
$$
We find the ADM mass per unit for this metric is 
$r_+/\sqrt{3} = 2a_0$ whereas the gravitational mass is $r_+/2\sqrt{3} = a_0$. 
The metric is however singular as $r\to r_+$. This singularity
is of a particularly unpleasant nature since it is both naked and has a
divergent volume element. To see that the singularity is naked, consider an
outgoing  radial null geodesic for which
$$
{dr\over dt} = \left ( 1 - {r_+\over r} \right ) ^{\sqrt{3}/2} 
\eqno(2.10)
$$
Therefore, for a null geodesic starting at $r_+$ at time $t_+$, the 
geodesic reaches
$r_+ + \delta r$ at time $t_+ + \delta t$ given by
$$
\delta t \propto  \delta r ^{1-\sqrt{3}/2}
\eqno (2.11)
$$
which is certainly finite.
However, since
$$
g_{tt} {dt\over d\lambda} =  \left ( 1 - {r_+\over r} \right )
^{1\over\sqrt{3}} {dt\over d\lambda} = {\rm constant}
\eqno (2.12)
$$
for $\lambda$ an affine parameter along the geodesic, any escaping photons are
infinitely redshifted.

This property in itself might appear to rule out these solutions in situations
of physical interest, however, an idealised (Nambu) string in four dimensions
has a ``naked singularity'' -- the conical singularity -- that is not even
null! It is only the gentle (integrable) nature of this singularity that
makes us tolerate it, as well as the fact that it has fairly convincingly 
been shown to be a good approximation to the real thing[13,14].
Since we are only looking for an exterior solution to a five-dimensional
cosmic string, we would hope that the core would somehow smooth out this
unpleasant behaviour, rather like the core of a four-dimensional cosmic string
smooths out the singular apex of the conical spacetime even in quite general
scenarios. Reversing the logic, one might hope that this singular behaviour
is simply the appropriate higher dimensional analogue of the conical
singularity.

To provide evidence for this claim, we will consider a fairly general core
model, whereby
$$
T^a_b = {1\over 4\pi}
{\rm diag} \;\; \{ E(r), E(r), - P_r(r), - P_\theta(r), - P_\theta(r)\}
.\eqno (2.13)
$$
The dominant energy condition will be assumed ($E\geq|P_*|$), and the
functions will be assumed to be effectively zero outside some finite core.
This assumption will replace more specific fall-off conditions, such as those
derived for the four-dimensional string[14], which would require a far more
detailed analysis than is appropriate here - we merely wish to show
that it is plausible to smooth out the singularity.  
We additionally make a weak field approximation, namely that
$$
\mu = \int_0^\infty \sqrt{-g} E(r) dr \ll 1
.\eqno (2.14)
$$
This approximation will mean that we can expand the equations of motion around
flat space, and hopefully derive a consistent solution. 

The flat space matter
equation, $T^{ab}_{~~,b}=0$, implies
$$
(r^2 P_r)' = 2r P_\theta
.\eqno (2.15)
$$
Since we are in five dimensions the Einstein equations read
$$
R^a_b = 8\pi ( T^a_b - {1\over 3} T \delta^a_b)
\eqno (2.16)
$$
and hence (2.2a-c) can be rearranged to give
$$
\eqalignno{
((A^4)'C^2)' &= {8 C^2 \over 3} ( E+P_r+2P_\theta) & (2.17a) \cr
\left ( {A'\over A} \right ) ^2 + \left ( {C'\over C}  \right ) ^2 
+ 4{A'C'\over AC} &= {1 + 2C^2P_r\over A^4C^2} & (2.17b) \cr
(A^4C^2)'' &= 2 + 4C^2(P_r+P_\theta) & (2.17c) \cr
}
$$
assuming $B=A^{-2}$ as before. We will integrate these equations out from the
core, making no assumption as to the asymptotic solution, although of course we
wish to show that the asymptotic form of the metric is the vacuum solution
(2.8), up to coordinate redefinition. We therefore impose the boundary
conditions at the core
$$
A=1 \;\;\; ; \;\;\; C\sim r \;\;\; {\rm as} \;\; r\to0
.\eqno (2.18)
$$
Note that, even if we had not assumed {\it a priori} 
that $g_{zz}=g_{tt}$, the form of
the energy-momentum tensor together with the above boundary conditions
(with $g_{zz}=1$ at $r=0$) would have necessitated $g_{zz}=g_{tt}$.

From (2.17a,c) we can see that the integration constants $a_0,~c_0$ in
(2.6) are given by
$$
a_0 = {2\over3}\mu \;\;\; ; \;\;\; c_0 = p -{4\over3}\mu
\eqno (2.19)
$$
where $p = \int_0^\infty r^2P_r$. To lowest order, (2.17c) can actually be
integrated up fully to give
$$
A^4C^2 = r^2 + 2r\int_0^r r^2P_r
.\eqno (2.20)
$$
The integrability constraint (2.17b) is then automatically
satisfied to first order,
and the first order solutions are given by
$$
\eqalignno{
A(r) &\simeq 1 - {2\over3} {\int_0^r r^2E\over r} + {2\over3}
\int_0^r r(E+P_r) & (2.21a) \cr
C(r) &\simeq r \left ( 1 - {4\over3} \int_0^r r(E+P_r) \right ) 
+ \int_0^r r^2 P_r + \int_0^r r^2E & (2.21b) \cr
}
$$
We can then see, again to first order in $\mu$, that the solution
asymptotes (2.8) up to a rescaling.

Obviously this does not prove that we can smooth out the singularity, but it
does at least provide encouragement that it might be possible to do so
by judicious choice of source. The main problem in finding a 
topological defect source for
the uncharged string is in the restriction to no long range interactions.
Since
the plane orthogonal to the vortex is now three-dimensional we are looking,
in the context of defects from spontaneous symmetry breaking, for a 
vacuum with non-trivial second homotopy group. If we spontaneously break
such a symmetry, we will not in general break it completely, for there is
still a residual U(1) symmetry around any point in the vacuum manifold.
This translates to a residual long range interaction and hence 
probable charge for the defect.  On the other hand, defects formed from
symmetry breaking are not the only types of soliton one could consider. It
is possible that the topology of the field space itself might be suitable for 
the formation of an extended field configuration with finite energy per unit
length. For example, the Skyrme model [15], which has localised 
finite energy field configurations
in four dimensions, might be a good candidate for the uncharged string. The 
flat space solution would satisfy our energy-momentum conditions, and appears
to be a promising source. However, one would have to check that
the extended solution in five dimensions did not exhibit any obvious
instabilities, as well as coupling the model to gravity - an involved
problem even in four dimensions[16].
Nonetheless, with suitable disclaimers\footnote{$^*$}{e.g. 
http://xxx.lanl.gov/legal/disclaimer.html},
we believe that the arguments given are indicative that the 
singularity is an artifact of the vacuum solution and that 
real strings will have non-singular cores.

Since we are in five-dimensions, let us discuss the Kaluza-Klein reduction
of our solution before proceeding to generalize it.
In a conventional Kaluza-Klein reduction, the five-dimensional metric
is written as
$$
ds^2 = - e^{4\sigma/\sqrt{3}} (dz + 2A_adx^a)^2 
+ e^{-2\sigma/\sqrt{3}} ~^4\!g_{ab}dx^adx^b
\eqno (2.22)
$$
which (after integration along z) yields an action
$$
\int d^4x \sqrt{-~^4\!g} \left [ {-R\over 16\pi} 
- {1\over 4} e^{2\sqrt{3}\sigma}
F_{ab}^2 + {1\over 8\pi} (\nabla \sigma)^2 \right ] 
\eqno (2.23)
$$
as the effective four-dimensional action. Since we have our five-dimensional
vacuum metric (2.8), we may read off:
$$
e^{4\sigma} = \left ( 1- {r_+\over r} \right ) 
\eqno(2.24)
$$
and
$$
ds_{_4}^2 = \left ( 1- {r_+\over r} \right )^{\sqrt{3}/2} dt^2
- \left ( 1- {r_+\over r} \right )^{-\sqrt{3}/2} dr^2 
- r^2 \left ( 1- {r_+\over r} \right )^{1-\sqrt{3}/2} d\Omega^2_{I\!I}
\eqno (2.25)
$$
this is again singular at $r=r_+$, although the volume element is slightly
better behaved. 

Such a four dimensional metric is not new, it is similar to the metrics of
Brans and Dicke[17] (see also Dicke[18])
which were derived in their original paper on 
Mach's principle and gravitation, although
Brans and Dicke considered the metrics in a conformally related
frame, equivalent to simply truncating the five-dimensional metric (2.8).
Additionally, 
Kaluza Klein theory corresponds to Brans-Dicke theory for $\omega=0$,
and therefore is technically outside the regime of the original 
Brans-Dicke results, however, provided one does not use Brans and Dicke's
values of their solution parameters in terms of $\omega$, which were
derived in the far field $\omega>3/2$ limit, the metrics can be seen
to agree. Indeed these solutions have been more recently considered in [19]
in a broader context, where it was argued that the naked singularities
corresponded to a non-trivial Parametrized Post Newtonian $\gamma$ 
parameter in the 
``Brans-Dicke'' frame. The PPN parameters
are used as a means of measuring how far a particular solution diverges
from the Einstein far field theory (see [20] for a review). Experimentally
$\gamma = 1 \pm .002$  for the Sun[21]. 
($\gamma=2$ for our truncated metric.) These solutions
were rejected in [19] as being physically unacceptable;
here however, 
we have obtained this metric by reduction of a five-dimensional object.
This metric is the metric of a bosonic Nambu string winding mode
in five dimensional Kaluza Klein theory and should certainly not
be ignored within the rationale of string theory. Moreover,
since we have argued that
the unpleasant singularity can most probably
be smoothed out by a core, this indicates
that such sources should not be regarded as physically unacceptable 
from the four-dimensional point of view, but rather, should be regarded
as being physically acceptable as dimensionally reduced solitons.
Additionally, since we could hardly claim that the Sun is a Nambu string
winding mode, the Viking data limits[21] on $\gamma$ do not really rule out
such solutions.
Now let us consider more general brane metrics.

\vskip 4mm

\noindent{\bf 3. General p-Branes.}
\vskip 2mm

Following (2.1), the general boost symmetric metric of a $p$-brane in
$N$-dimensions should have the form
$$
ds^2 = A^2 (dt^2 - dx_i dx^i) - B^2 dr^2 - C^2 d\Omega^2_{D-2}
\eqno (3.1)
$$
where $D=N-p$, and $i=1....p$ runs over the brane coordinates. 
The vacuum Einstein equations are
$$
\eqalignno{
R^0_0&= B^{-2} \left [ {A''\over A} - {A'B'\over AB}
+ p\left ({A'\over A} \right )^2 + 
(D-2) {A'C'\over AC} \right ] = 0 & (3.2a) \cr
R^r_r &=  B^{-2} \left [ (p+1){A''\over A} + (D-2){C''\over C}
- {B'\over B}\left ( (p+1){A'\over A} + (D-2){C'\over C} \right )
\right ] = 0 & (3.2b) \cr
R^\theta_\theta &=  B^{-2} \left [ {C''\over C}
+ {C'\over C} \left ( (D-3){C'\over C}+ (1+p){A'\over A} 
- {B'\over B} \right ) \right ]
- {(D-3)\over C^2} = 0. & (3.2c) \cr
}
$$
Since $B=A^{-2}$ was helpful in solving the five-dimensional cosmic
string, we will set $B=A^{-n}$ and search for a solution in a similar
fashion here.

As before, we can directly integrate (3.2a)  to obtain
$$
(A^{p+n+1})'C^{D-2} = a_0(p+n+1)
.\eqno (3.3)
$$
However, (3.2c) is no longer directly integrable. Instead, 
based on the intuition gleaned from the five dimensional string, we try
$$
A^{p+n+1} = \left ( 1 - \left ({r_+\over r} \right )^{D-3} \right )^m
\eqno (3.4)
$$
which gives immediately that
$$
C^{D-2} = r^{D-2} 
\left ( 1 - \left ({r_+\over r} \right )^{D-3} \right )^{1-m}
\eqno (3.5)
$$
from (3.3) and hence we see that our spacetime is asymptotically flat in 
the D-dimensions orthogonal to the brane. Substituting these forms
into (3.2b,c) give
two relations on $m$ and $n$ which, after some algebra, can be
solved to give
$$
\eqalignno{
n&= {p+1\over D-3} + {D-4\over D-3} \sqrt{(p+1)(D+p-2)\over(D-2)}
&(3.6a) \cr
m &= {D-4\over 2(D-3)} + {D-2\over2(D-3)} 
\sqrt{(p+1)(D-2)\over D+p-2} .& (3.6b) \cr
}
$$
Thus the general solution is given by (3.4), (3.5) and (3.6) with $B=A^{-n}$.
By setting $D=4$, $p=1$, we obtain the string of the previous section.
To illustrate the solution, we will consider two examples, the 5-brane in
ten dimensions and the $p$-branes in 4+$p$ dimensions.

\vskip 3mm
\noindent{\sl The 5-brane.}
\vskip 2mm

The 5-brane in 10 dimensions has $p=D=5$, therefore (3.6) gives
 $n=5$, $m=11/8$, 
and hence a metric:
$$
ds_{_5}^2 = \left ( 1 - {r_+^2\over r^2} \right ) ^{1\over4} (dt^2-dx^2_i)
-  \left ( 1 - {r_+^2\over r^2} \right ) ^{-{5\over4}}dr^2
- r^2  \left ( 1 - {r_+^2\over r^2} \right ) ^{-{1\over4}} d\Omega^2_{I\!I\!I}
\eqno (3.7)
$$
This again has a naked singularity at $r=r_+$, but should be smoothable
by a procedure analogous to that described in the previous section.

\vskip 3mm
\noindent{\sl The $p$-branes for $D=4$.}
\vskip 2mm

Another simpler family of solutions are the $p$-branes in $4+p$ dimensions
for which $n=1+p$ and $m = \sqrt{2(p+1)/(2+p)}$. Here
$$
ds^2 = \left ( 1 - {r_+\over r} \right ) ^{\sqrt{2}\over\sqrt{(p+1)(p+2)}}
(dt^2-dx^2_i) -  \left ( 1 - {r_+\over r} \right ) ^{\sqrt{2(p+1)}\over
\sqrt{(2+p)}}dr^2 - r^2  \left ( 1 - {r_+\over r} \right )^
{1 - \sqrt{2(p+1)\over(2+p)}} d\Omega^2_{I\!I}
\eqno (3.8)
$$
We may perform a Kaluza-Klein reduction by setting
$$
ds_N^2 = -\sigma^2 dx^2_i + \sigma^{-p} ds_4^2
\eqno (3.9)
$$
which yields the four-dimensional metric
$$
{\hbox{}}^4\!ds^2 =  
\left ( 1 - {r_+\over r} \right ) ^{\sqrt{p+2\over 2(p+1)}}dt^2
-  \left ( 1 - {r_+\over r} \right ) ^{-\sqrt{p+2\over 2(p+1)}}dr^2
- r^2 \left ( 1 - {r_+\over r} \right ) ^{1- \sqrt{p+2\over 2(p+1)}}
d\Omega^2_{I\!I}
\eqno (3.10)
$$
which gives a slightly different four-dimensional behaviour for each $p$,
and a PPN parameter $\gamma=(p+1)$ in the ``Brans-Dicke'' frame.

\vskip 4mm

\noindent{\bf 4. The Charged String.}

\vskip 2mm

We now wish to generalize the work of the previous two sections to include 
the effect of charge. We will begin by setting up the general formalism
before specializing to the five-dimensional string in this section. The 
next section will deal with the general branes. We
consider an action similar to that of 
Horowitz and Strominger[2], (see also [22] and references therein for a review 
of string solitons) namely an ``Einstein-Maxwell-Dilaton'' action 
with arbitrary dilaton coupling in N-dimensions:
$$
S = \int d^Nx \sqrt{-{\tilde g}} \left \{ e^{-2\phi} [ -{\tilde R}
-4({\tilde\nabla}\phi)^2]
(-)^{D-3} {2e^{2a\phi} F^2 \over (D-2)!} \right \}
\eqno (4.1)
$$
Here, $F$ is a ($D-2$)-form, $\phi$ the dilaton and $a$ the dilaton coupling.
However, rather than considering solving the field equations in the 
string frame, we will make a conformal transformation to the Einstein frame
where our analysis will follow more closely the previous two sections.

We therefore define
$$
g_{ab} = e^{-4\phi\over(N-2)} {\tilde g}_{ab}
\eqno (4.2)
$$
which gives us an action in Einstein form
$$
S = \int d^Nx \sqrt{-g} \left \{ - R
+ {4\over N-2} (\nabla \phi)^2 (-)^{D-3} 
{2F^2 e^{2\alpha\phi} \over (D-2)!} \right \}
\eqno (4.3)
$$
where 
$$
\alpha = a + {p+4-D\over N-2}
\eqno (4.4)
$$
gives the shifted dilaton coupling in the Einstein frame.
In this format, we can use the previous Einstein equations with the source
$$
T_{ab} = {4\over N-2} \nabla_a\phi\nabla_b\phi (-)^{D-3}
{2e^{2\alpha \phi}\over (D-3)!} F_{a...}F_b^{\ ...}
- g_{ab} \left [ {2\over N-2}(\nabla\phi)^2 (-)^{D-3} {F^2e^{2\alpha\phi}
\over(D-2)!} \right ] 
\eqno (4.5)
$$
We also have the dilaton
and electromagnetic equations of motion:
$$
\eqalignno{
\nabla_a \left [ e^{2\alpha \phi} F^{a...}\right ] &=0 & (4.6a) \cr
\bo \phi &= (-)^{D-3} {\alpha (N-2) \over 2(D-2)!} e^{2\alpha\phi}F^2
&(4.6b) \cr
}
$$

Following Horowitz and Strominger, we will first look for a magnetically
charged solution
$$
F = Q \epsilon_{D-2}
\eqno (4.7)
$$
where $\epsilon_{D-2}$ is the area form of a unit (D-2)-sphere.
We will also assume that $\phi = \phi(r)$. This form of $F$ then
automatically solves (4.6a). 
With these assumptions, and using the boost symmetric form of the metric
(3.1), the energy momentum tensor takes the form
$$
\eqalignno{
T^0_0 &= {2\over N-2} {\phi^{\prime2} \over B^2}
+ {Q^2 e^{2\alpha\phi}\over C^{2(D-2)}} & (4.8a) \cr
T^r_r &= - {2\over N-2} {\phi^{\prime2} \over B^2}
+ {Q^2 e^{2\alpha\phi}\over C^{2(D-2)}} & (4.8b) \cr
T^\theta_\theta &= {2\over N-2} {\phi^{\prime2} \over B^2}
- {Q^2 e^{2\alpha\phi}\over C^{2(D-2)}} & (4.8c) \cr
}
$$
and the dilaton equation
$$
\left ( {A^{p+1} C^{D-2} \phi' \over B} \right )' =
{(N-2)\alpha \over 2} {Q^2 e^{2\alpha\phi} A^{p+1}B\over C^{D-2}}
\eqno (4.9)
$$

Substituting $T^a_b-{T\over N-2}\delta^a_b$ as the source in the RHS of
(3.2) gives the Einstein equations
$$
\eqalignno
{
{ ( A^p A' C^{D-2}/B)'\over A^{p+1}C^{D-2} B} 
&= 2{(D-3)\over(N-2)} {Q^2e^{2\alpha\phi}\over C^{2(D-2)}} & (4.10a)\cr
{ ( A^{p+1} C^{D-3}C'/B)' \over A^{p+1}C^{D-2} B} - {D-3\over C^2}
&= -2{(p+1)\over(N-2)} {Q^2e^{2\alpha\phi}\over C^{2(D-2)}} & (4.10b)\cr
{(1+p)\over AB} \left ( {A'\over B}\right )'
+ {(D-2)\over CB} \left ( {C'\over B} \right )' &=
 -4 {\phi^{\prime 2}\over B^2(N-2)} + 
2{(D-3)\over(N-2)} {Q^2e^{2\alpha\phi}\over C^{2(D-2)}} & (4.10c)\cr
}
$$
Finally, we impose similar boundary conditions as in the uncharged branes,
namely that spacetime be asymptotically Minkowskian, $A,B\to1$, $C\sim r$
at infinity. Additionally,  for 
$\phi$ we will impose that $\phi\to0$ as $r\to\infty$ and that 
$\phi'$ is regular at the event horizon except possibly in some extremal
limit. The former boundary condition on $\phi$ is merely a choice for
algebraic simplicity, the latter boundary condition ensures that the weak
field limit $Q\ll M$ corresponds to a perturbation of the charge free brane
outside the event horizon.

These are obviously considerably more complicated than the uncharged
branes, so once more we will solve for the five-dimensional string
as a warm-up, before trying to tackle the full problem.
Therefore, we set $N=5$, $D=4$ and $p=1$, and as before, look for a
solution with set $B=A^{-2}$. The reason for choosing the Einstein frame 
now becomes slightly more apparent: in four dimensions, for black holes
in the string frame the time and
radial parts of the metric react differently to charge[23], but react
in the same way in the Einstein frame. Therefore, in five-dimensions,
we might
hope that in the Einstein frame we will have similar behaviour to the uncharged
branes and therefore that we can still use some of the results 
from the previous two sections.

With these substitutions, and some minor shuffling, the system of equations 
we must solve is given by
$$
\eqalignno
{
((A^4)'C^2)' &= {16\over 9\alpha} (A^4C^2\phi')' &(4.11a)\cr
(A^4(C^2)')' &= 2 - {16\over 9\alpha} (A^4C^2\phi')' &(4.11b)\cr
\left [ {(A^4C^2)'\over 2A^4C^2} \right ] ^2 - 3 \left ( {A'\over A}\right )^2
&= {1\over A^4C^2} + {2\over3}\phi^{\prime2} - {2\over3\alpha}
{(A^4C^2\phi')'\over A^4C^2} & (4.11c)\cr
(A^4C^2\phi')' &= {3\alpha\over2} {Q^2e^{2\alpha\phi}\over C^2} & (4.11d)\cr
}
$$
subject to the aforementioned boundary conditions.

We may proceed analogously to section two, (2.6a-c), to obtain
$$
\eqalignno{
(A^4)'C^2 &= {16\over 9\alpha} A^4C^2\phi' + 4a_0 & (4.12a) \cr
A^4(C^2)' &= 2r + 2c_0 - {16\over 9\alpha} A^4C^2\phi' & (4.12b) \cr
A^4C^2 &= r^2 + 2(c_0+2a_0)r + b_0 = (r-r_+)(r-r_-). & (4.12c) \cr
}
$$
However, the integrability condition (4.11c) does not give directly a
relation for $b_0$. Instead it gives an equation for $\phi$, which,
writing $f=A^4C^2\phi'$, can be seen to be a Ricatti equation for $f$:
$$
\eqalign{
f' &= { \left (\alpha + {8\over9\alpha}\right ) f^2 + 4a_0f -
{3\alpha\over2} (c_0^2 + a_0^2 +4a_0c_0-b_0) \over A^4C^2} \cr
&= {\beta f^2 +4a_0f + \gamma\over (r-r_+)(r-r_-)} \cr
}
\eqno (4.13)
$$
Now, regularity of $\phi'$ implies $\gamma=0$, and hence we have the
same roots $r_\pm$ as for the uncharged string. (4.13)  is then
readily integrated to give
$$
f = {4a_0f_\infty (r-r_+)^{2/\sqrt{3}} \over
(\beta f_\infty + 4a_0)(r-r_-)^{2/\sqrt{3}} - \beta f_\infty(r-r_+)^{2/\sqrt{3}}
}
\eqno (4.14)
$$
where
$$
f_\infty = \lim_{r\to\infty} A^4 C^2 f' = {1\over \beta}
\left (-2a_0 + \sqrt{4a_0^2 + 3\alpha \beta Q^2/2}\right )
.\eqno (4.15)
$$
using the equation of motion for $\phi$. We can then proceed with
the integration of the equations of motion (4.12a,b) to obtain in turn
$$
\eqalignno{
A e^{-{4\phi\over9\alpha}} &= \left ( {r-r_+\over r-r_-} \right )
^{1\over2\sqrt{3}} & (4.16a) \cr
C e^{8\phi\over9\alpha} &= (r-r_+)^{1-{2\over\sqrt{3}}}
(r-r_-)^{1+{2\over\sqrt{3}}} & (4.16b) \cr
e^{-\beta\phi} &=  1 + {\beta f_\infty \over 4a_0}
- {\beta f_\infty \over 4a_0} 
\left ( {r-r_+ \over r-r_-} \right ) ^{2\over\sqrt{3}}
.&(4.16c)\cr}
$$

Finally, we use the remaining coordinate freedom (choice of r-origin) to set
$ r_- = \beta f_\infty $
which implies
$$
Q^2 = {r_+r_- \over 4\alpha\beta},
\eqno (4.17)
$$
and gives the five dimensional boost-symmetric dilatonic charged string
metric
$$
\eqalign{
g_{tt} &= \left [ 
 1 + {\sqrt{3}r_-\over 2(r_+-r_-)} - {\sqrt{3}r_-\over 2(r_+-r_-)}
\left ( {r-r_+ \over r-r_-} \right ) ^{2\over\sqrt{3}}
\right ] ^{{-8\over9\alpha\beta}} \left ( {r-r_+ \over r-r_-} \right ) 
^{1\over\sqrt{3}} \cr
g_{\theta\theta} &= \left [ 
 1 + {\sqrt{3}r_-\over 2(r_+-r_-)} - {\sqrt{3}r_-\over 2(r_+-r_-)}
\left ( {r-r_+ \over r-r_-} \right ) ^{2\over\sqrt{3}}
\right ] ^{{16\over9\alpha\beta}} \hskip -5mm 
(r-r_-)^{1 + {2\over\sqrt{3}}}
(r-r_+)^{1 - {2\over\sqrt{3}}} \cr
g_{zz} &= g_{tt} \;\;\; ; \;\;\; g_{rr} = (g_{tt})^{-2} \;\;\; ; \;\;\;
g_{\phi\phi} = \sin^2\theta g_{\theta\theta} \cr
}\eqno (4.18)
$$

It is interesting to examine this metric for a couple of special cases:

\vskip 2mm

\noindent{i) $\alpha=0$}

\vskip 1mm

For $\alpha=0$ the dilaton completely decouples from electromagnetism, and
what we are left with is Einstein-Maxwell gravity in five dimensions.
Noting that $\alpha\beta=8/9$, we see that the charged string has metric
$$
ds^2 = A^2 (dt^2-dz^2) - A^{-4}[dr^2 + (r-r_+)(r-r_-)d\Omega^2_{I\!I}]
\eqno (4.19)
$$
where
$$
A^2 = \left [ 1 + {\sqrt{3} r_-\over 2(r_+-r_-)} \left (
1 - \left ( {r-r_+\over r-r_-}\right )^{2/\sqrt{3}} \right ) \right ] ^{-1}
\left ( {r-r_+\over r-r_-}\right )^{1/\sqrt{3}}.
\eqno (4.20)
$$
We may compare this with the metric obtained by ignoring boost symmetry, and
merely extending the magnetically charged Reissner Nordstrom solution by
adding a flat direction:
$$
ds^2 = \left ( 1 - {r_+\over r} \right ) \left ( 1 - {r_-\over r} \right )
dt^2 - dz^2 -  \left ( 1 - {r_+\over r} \right ) ^{-1}
\left ( 1 - {r_-\over r} \right )^{-1}dr^2 - r^2 d\Omega^2_{I\!I}
\eqno (4.21)
$$
Clearly these two metrics are rather different! However, if one were to 
consider the metric as arising from, say an SU(2) monopole in five dimensions,
then the boost symmetry of the energy-momentum tensor would require the
metric (4.20) rather than (4.21). The only remaining question is whether
the equation of motion would integrate out to this asymptotic form.

It is beyond the scope of this paper to
analyse the non-linear field equations resulting from such a substitution,
however, by referring to the first order corrections to the metric (2.21)
obtained for a rather general energy-momentum tensor, we can see that
provided the defect is weakly gravitating in the sense of (2.14), the
$1/r^2$ fall-off typical of a charged source should not obstruct the
integrals in (2.21) remaining finite, and the indications are that the
metric will indeed integrate out to the asymptotic form (4.25).
We can make some observations on the BPS limit, in this limit, the 
potential of the Higgs field vanishes and the equations of motion become
first order in the absence of gravity. The now massless scalar acquires
a long range fall-off which exactly counterbalances the electromagnetic
fall-off. The energy-momentum tensor in flat space, as expected, has no
non-zero components orthogonal to the monopole worldsheet.
Without reference to the specific fields, one can see that coupling in
gravity destroys this balance, as it does in four-dimensions[24,25]. 
This can be understood as a consequence of the attractive nature of 
gravity requiring some radial pressure to support against collapse. 

The extremal limit of (4.19) takes the form
$$
\left ( 1- {r_+\over r} \right ) (dt^2-dz^2) - \left (
1- {r_+\over r} \right )^{-2} dr^2 - r^2 d\Omega^2_{I\!I}
\eqno (4.22)
$$
which does not appear to be singular as $r\to r_+$. However, for $r<r_+$
something curious occurs. Normally, it is the time and radial coordinates
that swap roles across the event horizon, $r$ becoming timelike, and $t$
spacelike. However, in the above metric, it appears to be $t$ and $z$ that
are trading places, i.e., the worldsheet coordinates.

\vskip 2mm

\noindent{ii) $\alpha=-2/3$}

\vskip 1mm

This special case corresponds to $a=-1$, or a dilaton coupling usually
associated with low energy string gravity. For $\alpha = -2/3$,
$\beta = -2$, and therefore we have
$$
e^{2\phi} = 
\left [
 1 + {\sqrt{3}r_-\over 2(r_+-r_-)} - {\sqrt{3}r_-\over 2(r_+-r_-)}
\left ( {r-r_+ \over r-r_-} \right ) ^{2\over\sqrt{3}}
\right ]
\eqno(4.23)
$$
By noting that $8\over9\alpha\beta$=${4\over3}$ we have the exponents for the
metric in (4.18), however, it is more enlightening to transform back to the
string frame, multiplying by a conformal factor of $e^{4\phi/3}$ to obtain
the metric in the string frame:
$$
\eqalign{
d{\tilde s}^2 &= \left ( {r-r_+\over r-r_-} \right ) ^{1\over\sqrt{3}}
(dt^2-dz^2) - e^{4\phi} \left ( {r-r_+\over r-r_-} \right ) ^{-4\over\sqrt{3}}
[dr^2 + (r-r_+)(r-r_-) d\Omega^2_{I\!I}] \cr
&\to dt^2-dz^2 - \left( 1-{r_+\over r} \right )^{-2} dr^2
-r^2 d\Omega^2_{I\!I} \hskip 1cm {\rm as} \;\; {r_-\to r_+} \cr
}
\eqno (4.24)
$$
This will be recognised as the truncated extremal limit of the
Horowitz-Strominger 6-brane in 10 dimensions.

Now that we have an idea of the steps involved, we 
will tackle the general brane.

\vskip 4mm

\noindent{\bf 5. General Charged Branes.}

\vskip 2mm

We would now like to solve for the general charged $p$-brane. As before, we
will begin by assuming that the brane is magnetically charged. In the next
section we will show how to derive an electrically charged brane via a
duality transformation.

Clearly (4.9) and (4.10a) imply
$$
{A^pA'C^{D-2} \over B} = a_0 + {4(D-3)\over \alpha(N-2)^2}\phi'
{A^{p+1} C^{D-2} \over B}
\eqno (5.1)
$$
and we might expect that, just as the five dimensional string metric maintains
the same powers of $(1-r_+/r)$, the general brane might be similarly related to
the uncharged brane, and look for a solution
$$
A^{p+n+1} = \left ( {r^{D-3} - r_+^{D-3} \over r^{D-3} - r_-^{D-3}}\right )^m
\exp \left \{ {4(D-3)(p+n+1)\phi\over \alpha(N-2)^2} \right \}
\eqno (5.2)
$$
with $m$ given by (3.6b). 
However, the substitution $B=A^{-n}$ rapidly leads to
problems! Therefore, we relax the idea that the time and radial
components of the metric react the same way to  $r_-$, and 
instead try a substitution of the form
$$
B= A^{-n} e^{\gamma\phi} \left ( 1 - \left ({r_-\over r}\right )^{D-3} \right
)^s \eqno (5.3)
$$
where $n$ is given by (3.6a). 
Substituting these guesses into (5.1) gives
$$
C^{D-2} = r^{D-2} \left ( 1 - ({r_+\over r} )^{D-3} \right )^{1-m}
\!\!\left ( 1 - ({r_-\over r})^{D-3} \right )^{1+m+s} \!\!\!\!\!
\exp \left \{  \gamma\phi - {4(p+n+1)(D-3)\phi\over\alpha(N-2)^2} 
\right \}
\eqno (5.4)
$$
It is then straightforward, if lengthy, to show that these functions satisfy
(4.10b) provided
$$
\eqalignno{
s &= - {D-4\over D-3} & (5.5a) \cr
\gamma &= {4\over \alpha(N-2)^2} \left [ n(D-3) - (p+1) \right ] & (5.5b) \cr
}
$$
Note that for $D=4$, $\gamma=s=0$ in agreement with the solution already
derived for the five dimensional string.

Finally, writing
$$
g = {A^{p+1}C^{D-2} \over B} \;\;\;\;\; ; \;\;\;\; f = g\phi'
\eqno (5.6)
$$
and substituting
(5.2-4) into (4.10c) yields, as in the case of the
five dimensional string, a Ricatti
equation for  $f$ 
$$
gf' = \beta f^2 + l(D-3)(r_+^{D-3} - r_-^{D-3})f
\eqno (5.7)
$$
where
$$
\beta = \alpha + {4(D-3)(p+1)\over \alpha(N-2)^2}
\eqno (5.8)
$$
generalises the $\beta$ of equation (4.13), and
$$
l = {2m(p+1)\over p+n+1} = \sqrt{(p+1)(D-2)\over(N-2)}
.\eqno (5.9)
$$
Additionally, (4.9) gives
$$
gf' = {(N-2)\alpha Q^2\over2} e^{2\beta\phi} \left ( {r^{D-3} - r^{D-3}_+\over
r^{D-3} - r^{D-3}_-}\right )^l
\eqno (5.10)
$$
One can then verify that
$$
e^{-\beta\phi} =  1 + {r_-^{D-3} \over l(r^{D-3}_+ - r^{D-3}_-)} \left [
1 - \left ( {r^{D-3} - r^{D-3}_+\over r^{D-3} - r^{D-3}_-}\right )^l
\right ]
\eqno (5.11)
$$
is the solution satisfying these two equations
and the boundary conditions, with the charge, $Q$, being given by
$$
Q^2 = {2(D-3)^2 r^{D-3}_- \over \alpha\beta(N-2) }\left ( l (r^{D-3}_+ -
r^{D-3}_-) + r^{D-3}_- \right ) .
\eqno (5.12)
$$

To summarize, the metric of the charged $p$-brane is given by
$$
\eqalign{
ds^2 =\;\;\; &e^{8(D-3)\phi\over\alpha(N-2)^2} \left ( {r^{D-3}
- r^{D-3}_+\over r^{D-3} - r^{D-3}_-}\right )^{\sqrt{D-2\over(p+1)(N-2)}}
 (dt^2 - dx_i^2) \cr
&- e^{-8(p+1)\phi\over\alpha(N-2)^2} \left ( {r^{D-3} - r^{D-3}_+\over
r^{D-3} - r^{D-3}_-}\right )^{-{1\over D-3}[\sqrt{(p+1)(D-2)\over(N-2)}+(D-4)]}
\left ( 1 - ({r_-\over r})^{D-3}
\right ) ^{-2(D-4)\over D-3} dr^2 \cr
& - e^{-8(p+1)\phi\over\alpha(N-2)^2}\left ( {r^{D-3} - r^{D-3}_+\over
r^{D-3} - r^{D-3}_-}\right )^{{1\over D-3}[1-\sqrt{(p+1)(D-2)\over(N-2)}]}
\left ( 1 - ({r_-\over r})^{D-3}\right ) ^{2\over (D-3)}
r^2 d\Omega^2_{D-2}\cr
}
\eqno (5.13)
$$
with the dilaton given by (5.11). 
An alternative form of the metric which is also useful is given by making
the coordinate change
$$
\eqalign{
y^{D-3} &= r^{D-3} - r_-^{D-3} \cr
y_0^{D-3} &= r_+^{D-3} - r_-^{D-3} \cr
}
\eqno(5.14)
$$
in which case the metric has the form
$$
\eqalign{
ds^2 &= e^{8(D-3)\phi\over\alpha(N-2)^2} \left ( 1 -
( {y_0\over y} )^{D-3} \right ) ^{\sqrt{D-2\over(p+1)(N-2)}}
 (dt^2 - dx_i^2) \cr
&- e^{-8(p+1)\phi\over\alpha(N-2)^2} \left ( 1 - 
( {y_0\over y} )^{D-3} \right )
^{-{1\over D-3}[\sqrt{(p+1)(D-2)\over(N-2)}+(D-4)]}
\left [ dy^2 + y^2 \left ( 1 - 
( {y_0\over y} )^{D-3} \right )d\Omega^2_{D-2}\right ] \cr
}
\eqno(5.15)
$$
Note that the extremal limit of this solution now corresponds to $y_0=0$

It is obviously interesting to contrast these solutions with those of Horowitz
and Strominger [2]. Recall that the magnetically charged HS solutions have the
form
$$
e^{-2\phi} = \left [ 1 - \left ( {r_-\over r}\right ) ^{D-3} \right ]^{-2\over
\beta} \eqno (5.16)
$$
$$
\eqalign{
d{\tilde s}^2 = \;\; &\left [ 1 - \left ( {r_+\over r}\right ) ^{D-3} \right ]
\left [1 - \left ( {r_-\over r}\right ) ^{D-3} \right ]^{\gamma_x-1} dt^2
- \left [ 1 - \left ( {r_-\over r}\right ) ^{D-3} \right ]^{\gamma_x} dx_i^2 \cr
&-\left [ 1 - \left ( {r_+\over r}\right ) ^{D-3} \right ]^{-1}
\left [ 1 - \left ( {r_-\over r}\right ) ^{D-3} \right ]^{\gamma_r} dr^2
-r^2 \left [ 1 - \left ( {r_-\over r}\right ) ^{D-3} \right ]^{\gamma_r +1}
d\Omega^2_{D-2} }
\eqno (5.17)
$$
where the exponents are given by
$$
\textstyle{ \gamma_x = {4\alpha + D -3 \over 8\alpha\beta} \;\;\; ;\;\;\;
\gamma_r = {1\over2\alpha\beta} \left ( \alpha + {(D-11)\over 4} \right ) -
{(D-5)\over (D-3)} 
}
\eqno (5.18)
$$
and
$$
Q^2 = {(D-3)^2 (r_+r_-)^{D-3} \over 4\alpha\beta}
\eqno (5.19)
$$
in terms of our constants $\alpha$ and $\beta$ with $N=10$.

Obviously these solutions are rather different, as is expected, since the
HS solutions do not have boost symmetry, however, since the HS solutions 
{\it are} boost symmetric in their extremal limit, we would expect that
they would agree with ours. Indeed, 
the dilaton solutions, (5.11) and (5.16), 
do  give the same extremal solution,
with (5.12) and (5.19) agreeing upon the charge for that solution. 
To compare the metrics, we must conformally transform (5.17) via (4.2),
remembering that $N=10$, before comparing with (5.13),which does indeed
give the same extremal limit.

In general,  the extremal limit of our solution is 
$$
\eqalign{
ds^2_e &= \left ( 1 - ({r_+\over r})^{D-3} \right )
^{8(D-3)\over\alpha\beta(N-2)^2} (dt^2-dx^2_i) - 
\left ( 1 - ({r_+\over r})^{D-3} \right )^{ {-8(p+1)\over \alpha\beta(N-2)^2}
-{2(D-4)\over D-3}} dr^2 \cr
& \;\;\; - r^2 \left ( 1 - ({r_+\over r})^{D-3} \right )^{{2\over (D-3)}
-{8(p+1)\over\alpha\beta(N-2)^2}} d\Omega^2_{D-2}\cr
&= \left ( 1+ ({r_+\over y})^{D-3} \right )
^{-8(D-3)\over\alpha\beta(N-2)^2} (dt^2-dx^2_i) -\left ( 1+({r_+\over y})
^{D-3} \right )^{
{8(p+1)\over\alpha\beta(N-2)^2}} \left [ dy^2 + y^2 d\Omega^2_{D-2}
\right ]
}
\eqno (5.20)
 $$
These are the metrics for extremally magnetically charged $p$-branes
in arbitrary dimension. Some of these metrics have already been derived
by Duff and Lu[26], who found these solutions for specific values of
$\alpha$, however, note here that we have no restriction on the value of
$\alpha$.

We can therefore be reasonably confident that our solutions do represent
true boost-symmetric $p$-branes.
Clearly they lack the relative simplicity of the other GMHS
solutions, they are rather messy and still suffer from singularities.
Whether or not they can be convincingly shown to be the far field limit
of some soliton or topological defect requires not only a choice of model,
but also a decision on how to couple that matter to the string dilaton -
a problem we will not address in this paper. However, we do believe that
in principle, there is no obstruction to painting a $p$-brane core onto
the spacetime to smooth out the singularity.

As before, we would like to consider some specific examples. 
Let $a=-1$, then $\alpha = {-2(D-3)\over (N-2)}$, and 
it turns out that $\beta = -2$ for all $N$ and $D$. We also prefer to
invert the conformal transformation (4.2) to quote results in the
possibly more familiar string frame. The dilaton solution is
given by (5.11) with $\beta=-2$, and the metric in the string frame
is 
$$
\eqalign{
&d{\tilde s}^2 = \left ( {r^{D-3} - r^{D-3}_+ \over r^{D-3} - r^{D-3}_-
}\right ) ^{l\over p+1} (dt^2 - dx^2_i)- 
 e^{4\phi\over(D-3)}
{\textstyle \left ( {r^{D-3} - r^{D-3}_+ \over r^{D-3} - r^{D-3}_-} \right )
^{-(l+D-4)\over D-3}}\times \cr
&\left [ {\textstyle
 \left ( 1 - ({r_-\over r})^{^{D-3}} \right ) ^
{-2(D-4)\over(D-3)}} dr^2 + {\textstyle 
r^2 \left ({r^{D-3} - r^{D-3}_+ \over r^{D-3} - r^{D-3}_-}
\right ) \left ( 1 - ({r_-\over r})^{^{D-3}}\right )^{2\over D-3}
 } d\Omega^2_{_{D-2}}
 \right ]
\cr
}
\eqno (5.21)
$$
For $D=5$, $N=10$ we get the 5-brane: $n=5$, $m=11/8$, $l=3/2$, and
$$
\eqalign{
d{\tilde s}^2 = &\left({r^2 - r_+^2 \over r^2 - r_-^2}\right )^{1/4}
(dt^2 - dx_i^2) \cr
&- e^{2\phi} 
\left({r^2 - r_+^2 \over r^2 - r_-^2}\right )^{-5/4} (1-{r_-^2\over r^2})
^{-1} \left [ dr^2 + r^2 \left ( 1 - {r_+^2\over r^2}\right )
\left ( 1 - {r_-^2\over r^2}\right ) d\Omega^2_{I\!I\!I}\right ]
\cr}
\eqno (5.22)
$$
with dilaton
$$
e^{2\phi} = 1 + {2 r_-^2 \over 3(r_+^2 - r_-^2)} \left [
1 - \left ( {r^2-r_+^2 \over r^2 - r_-^2} \right )^{3\over 2} \right ]
\eqno (5.23)
$$
This is the metric and dilaton of the non-supersymmetric 5-brane.

For $D=4$ and $N=10$, we have the 6-brane of heterotic string gravity:
$$
d{\tilde s}^2 = \left ( {r-r_+\over r-r_-}\right )^{1\over 2\sqrt{7}}
(dt^2 - dx^2_i) - e^{4\phi} \left ( {r-r_+\over r-r_-}\right )^
{-\sqrt{7}\over 2}
\left [ dr^2 + (r-r_+)(r-r_-) d\Omega^2_{I\!I}\right ]
\eqno (5.24)
$$
with dilaton
$$
e^{2\phi} = 1 + {2r_-\over \sqrt{7}(r_+-r_-)}\left [ 1 -
\left ( {r-r_+\over r-r_-}\right )^{\sqrt{7}\over2}\right ]
\eqno (5.25)
$$
\vskip 4mm

\noindent{\bf 6. Dual and Self-Dual Branes.}

\vskip 2mm

The solutions of the previous two sections were all magnetically
charged, 
obviously we would like to find electrically charged
solutions. Fortunately, a generalisation of the duality transformation of
Horowitz and Strominger allows us to derive these solutions.
Define
$$
K = e^{2\alpha\phi}\ast F
\eqno (6.1)
$$
where $\ast$ is the Hodge dual,
then (4.6a) is equivalent to the statement that $K$ is a closed
$(N-D+2)$-form. We can then see that
$$
K^2 = (-)^{N-1} e^{4\alpha\phi}F^2
\eqno (6.2)
$$
Then, provided we write $D' = N+4-D$ and $\alpha'=-\alpha$,
the energy-momentum tensor (4.8), the dilaton equation (4.9), and
hence the Einstein equations (4.10) are  invariant
under the operation $\{F,D,\alpha\}\to\{K,D',\alpha'\}$.
Thus to get an electrically charged solution, we take the magnetic
solution for $N+4-D$ and $-\alpha$, and dualise the magnetic form
according to (6.1).

For example, if we dualize the 2-form $F_{ab}$ of heterotic string gravity,
then we get a 0-brane, or black hole, in ten dimensions:
$$
d{\tilde s}^2 = \left ( 1 - ({r_+\over r})^7 \right ) \left (
1 - ({r_-\over r})^7 \right ) dt^2 - {dr^2 \over 
\left ( 1 - ({r_+\over r})^7 \right ) \left (1 - ({r_-\over r})^7 \right )
^{5/7}} - r^2 \left ( 1 - ({r_-\over r})^7 \right )^{2/7} d\Omega^2_8
\eqno (6.3)
$$
with dilaton
$$
e^{2\phi} = 1 - ({r_-\over r})^7
\eqno (6.4)
$$
which, not surprisingly, is identical to the solution in [2].

However, given the motivation of string theory in our choice of action
(4.1) it is interesting to ask what a string or 1-brane solution in ten
dimensions will look like, since it is believed that the extremal
limit[27] of such an electrically charged solution actually {\it is} a
fundamental string. An electrically charged 1-brane has $D'=9$, or
$D=5$, and hence corresponds to the axion field, $H_{abc}$, carrying
``electric'' charge. We take $a=-1$ in the original action (4.1) which 
corresponds to $\alpha=-1/2$. Therefore, to get our family of strings,
we dualize the magnetic $D=9, \alpha=1/2$ solutions. Choosing the 
$y$-coordinate as defined in (5.14), we may read off our solution
from (5.15), (5.11) as:
$$
ds^2 = \Lambda (dt^2 - dx^2_i) - \Lambda^{-1/3} \left [ dy^2
\left (1 - ({y_0\over y})^6  \right )^{-5/6} + y^2
\left ( 1 - ({y_0\over y})^6  \right )^{1/6} 
d\Omega^2_7 \right ]
\eqno(6.5)
$$
where
$$
\eqalign{
\Lambda &= e^{3\phi/2} \left ( 1 - ({y_0\over y})^6  \right ) ^{\sqrt{7}/4}
\cr
e^{-2\phi} &= 1 + {2r_-^6 \over \sqrt{7} y_0^6}
\left [ 1 - \left ( 1 - ({y_0\over y})^6  \right ) ^{\sqrt{7}/2}
\right ] \cr}
\eqno(6.6)
$$
and the axion field is given by:
$$
H_{abc} = Qe^{3\phi} *\epsilon_7
\eqno(6.7)
$$
in the Einstein frame. To get to the string frame
$$
\eqalign{
{\tilde g}_{ab} = e^{\phi/2} g_{ab} \cr
{\tilde H}_{abc} = e^{-\phi} H_{abc} \cr
}
\eqno(6.8)
$$
which gives for the metric
$$
\eqalign{
ds^2 = &e^{2\phi} \left (1 - ({y_0\over y})^6  \right )^{\sqrt{7}/4}
(dt^2-dx^2_i) \cr &- \left (1 - ({y_0\over y})^6  \right )^{-\sqrt{7}/12}
 \left [ dy^2
\left (1 - ({y_0\over y})^6  \right )^{-5/6} + y^2
\left ( 1 - ({y_0\over y})^6  \right )^{1/6}
d\Omega^2_7 \right ]
\cr}
\eqno (6.9)
$$
It can be seen that the extremal limit is indeed the fundamental string
of [27].

So far, we have been examining either electrically or magnetically charged
branes, but if the spacetime dimension, $N$, is even, then it is possible
that the charge be a self-dual (or anti-self-dual) $N/2$-form. For such
sources,
$$
F^2 = (-)^{N/2} F^2
\eqno (6.10)
$$
and therefore if $N=4j+2$ for some $j\in$ \zed, then $F^2$ vanishes, and
the dilaton decouples from the equations of motion. Taking F to be the
self dual form
$$
F = {Q\over\sqrt{2}} [ \epsilon_{N/2} + \ast  \epsilon_{N/2}]
\eqno (6.11)
$$
gives the energy-momentum tensor (4.8) with $\phi$ set to zero. Hence the
self-dual ${N\over2}-2$-branes are given by the $\alpha=0$ solutions of section
five. (In reality, $\alpha=a$.)
$$
\eqalign{
& ds^2 =  \Xi (dt^2 - dx_i^2) \cr
&-\Xi^{-1} \left ( 1 - ({r_+\over r})^{N-2\over2} \right )
^{2\over N-2}
\left ( 1 - ({r_-\over r})^{N-2\over2} \right )
^{2\over N-2}
\left [ {dr^2 \over  \left ( 1 - ({r_+\over r})^{N-2\over2} 
\right )\left ( 1 - ({r_-\over r})^{N-2\over2} \right )
} + r^2 d\Omega^2_{N/2}\right]
\cr }
\eqno (6.12)
$$
where
$$
\Xi = \left ( { r^{N-2\over2} - r_+^{N-2\over2}\over r^{N-2\over2}
-r_-^{N-2\over2}} \right ) ^ {\sqrt{N}\over N-2} \left [ 1
+ {2r_-^{N-2\over2} \over \sqrt{N} (r_+^{N-2\over2}-r_-^{N-2\over2})}
\left ( 1 - \left ( { r^{N-2\over2} - r_+^{N-2\over2}\over r^{N-2\over2}
-r_-^{N-2\over2}} \right ) ^ {\sqrt{N}\over2} \right ) \right ] ^{-4\over N-2}
\eqno(6.13)
$$
\vskip 4mm

\noindent{\bf 7. Discussion.}

\vskip 2mm

We have derived the metrics for a boost symmetric $p$-branes
in an arbitrary number of spacetime dimensions. We use an 
Einstein-Maxwell-Dilaton action for arbitrary values of the dilaton
coupling and arbitrary values of the charge carried by the `Maxwell'
field. Previsouly, the only boost symmetric metrics considered
were those with maximal, or extremal, charge. We find that the
exact solutions generically have naked singularities
at the brane core, however, we believe that this is an aspect of
the idealised solution and that is one were to consider
instead the branes as arising as defects in some spontaneously
broken gauge theory, this unpleasant behaviour would be smoothed out
by the internal structure of the defect. We have shown in principle
that there is no obstruction to doing this.

An alternative to placing a source at the $p$-brane core is instead to
regard these solutions as leading order solutions to string gravity, and 
therefore the singularity might get smoothed out by higher order
stringy effects 
(see e.g.~[28,29]).
In other words, string theory might act as a cosmic
censor. Since the bulk of the solutions presented here are not extremal,
and hence not supersymmetric or BPS states, they would certainly
be expected to acquire higher order or loop corrections in a fully consistent
expansion. However, we should point out that while the solutions 
presented here display the same strong-weak/electric-magnetic 
duality present in previous results, the main
difference is that the singularity actually sits at $r_+$, rather than
lying inside some event horizon. The choice of  boundary
conditions for the dilaton means that the dilaton is always finite, whether
the brane is magnetically or electrically charged, at this
point - except in the extremal limit.  In other words, if we identify 
$e^\phi$ as our string loop coupling, then the singularities always occur in a
weakly, or at least not strongly, coupled r\'egime. Therefore we can only
appeal to O($\alpha'$) arguments[28]. Whether or not such a process
would smooth out the singularity is an open question, indeed, whether one
should smooth out singularities in general is an open question[30]! However,
having a naked singularity is rather disturbing for the more
deterministic!

One interesting facet of the HS solutions is that they are unstable[3]
except possibly in their extremal limit. This is interesting precisely because
it is only in the extremal limit that these solutions are boost symmetric.
It would be useful to know if the solutions presented here are
stable. Indeed, since we are arguing that they correspond to the far field 
of some topological defect, it would be very surprising if they did 
exhibit a linear instability. If they are stable, then they
could resolve the question of the endpoint of the instability in [3].
One of the curiosities of the instability was that it appeared to 
lead to fragmentation of the event horizon, this in turn would lead to
a violation of cosmic censorship. While there was no obstruction, other
than the mythical cosmic censor, to this occurring for the uncharged solutions,
it did seem that the magnetically charged solutions should not be able to
break, for the simple reason that the charge is topological, and there 
would be nowhere for it to go. It has been suggested that there was
some endpoint other than fragmentation for this instability\footnote{$^\dagger$}
{I would like to thank G.Gibbons and G.Horowitz for discussions on this
point.}. Perhaps
the $p$-branes presented here represent that endpoint?

Of course, as we remarked in the introduction, topology does not guarantee
stability, for there might be a higher dimensional generalisation
of the C-metric which would provide an instanton for the decay
of some or all of the $p$-branes. Briefly, in four-dimensions, the
C-metric corresponds to two black holes accelerating apart, attached to
infinity by conical deficits responsible for their acceleration. It was 
shown in [10] that these deficits could be smoothly replaced by a
Nielsen-Olesen vortex, and hence that an otherwise stable
topological defect could decay as suggested in [7-9]. Suppose one
generalises the C-metrics to higher dimensions, then one has a process
for the decay of an idealized $p$-brane in $N$-dimensions. Given 
such an instanton, it seems likely that at least some of the solutions
found here would be susceptible to just such a non-perturbative
decay process. It is notable that the decay of strings in four-dimensions
is insensitive to whether the vortex is sitting at its Bogomolnyi point or not.
Therefore, if there is a non-perturbative decay process which is indifferent
to BPS states, it is possible that some of these states might
also be unstable. Clearly, if certain extremal
states are considered to be of importance in stringy duality arguments,
testing their resilience to non-perturbative decay processes is crucial. 
We have provided a first step in that direction here by generalizing 
the conical singularity to higher dimensions.
It would be amusing, although admittedly unlikely, if the extremal
black strings were unstable to non-perturbative topology changing processes.
Often, supersymmetry is invoked for protection against instability, for
example, in the decay of the Kaluza-Klein vacuum[31]. However, in an intriguing
recent paper concerning the decay of Kaluza-Klein magnetic
universes[32],  it was shown that while supersymmetry did protect against a 
decay of the Witten type[31], it did not protect against decay of the
black hole pair nucleation type. Such a result does not bode well for
the decay of strings in higher dimensions.

\vskip 4mm

\noindent{\bf Acknowledgements.}

\vskip 2mm

I would like to thank Peter Bowcock and Jeff Harvey for useful
comments and suggestions. This work was supported by the Royal Society.

\vskip 4mm

\noindent{\bf References.}
\vskip 2mm

\nref
G.W.Gibbons and K.I.Maeda, \npb 298 741 1988.

\nref
G.Horowitz and A.Strominger, \npb 360 197 1991.

\nref
R.Gregory and R.Laflamme, \prl 70 2837 1993. [hep-th/9301052] 
\npb 428 399 1994. [hep-th/9404071]

\nref
R.Gregory and R.Laflamme, \prd 51 305 1995. [hep-th/9410050]

\nref
M.Hindmarsh and T.W.B.Kibble, \rprog 58 477 1995. [hep-ph/9411342]

\nref
R.H.Brandenberger, {\it Modern Cosmology and Structure Formation}
astro-ph/9412049.

\nref
S.W. Hawking and S.F. Ross, \prl 75 3382 1995. [gr-qc/9506020]
 
\nref
R. Emparan, \prl 75 3386 1995. [gr-qc/9506025]
 
\nref
D. Eardley, G. Horowitz, D. Kastor and J. Traschen, 
\prl 75 3390 1995. [gr-qc/9506041]
 
\nref 
R.Gregory and M.Hindmarsh, \prd 52 5598 1995. [gr-qc/9506054] 

\nref
E.B.Bogomolnyi, {\it Yad.~Fiz.} {\bf24} 861 (1976)[{\it
Sov.~J.~Nucl.~Phys.~{\bf 24}} 449 (1976)]
 
\nref
W. Kinnersley and M. Walker, \prd 2 1359 1970.
 
\nref
D.Garfinkle, \prd 32 1323 1985.

\nref
R.Gregory, \prl 59 740 1987.

\nref 
T.H.R.Skyrme, \prsa 260 127 1961.

\nref
P.Bizon and T.Chmaj, \plb 297 55 1992.
 
\nref
C.Brans and R.H.Dicke, \pr 124 925 1961.

\nref
C.H.Brans, \pr 125 2194 1962.

\nref
S.Kalyana-Rama, {\it Singularities in low energy D=4 heterotic string
and Brans-Dicke theories}, hep-th 9309046.

\nref
C.M.Will, \book Theory and Experiment in Gravitational Physics 
[[Cambridge University Press, 1981]]

\nref
R.D.Reasenberg et.al., \apj 234 L219 1979.

\nref
M.J.Duff, R.R.Khuri and J.X.Lu, \phrep 259 213 1995. [hep-th/9412184]

\nref 
D.Garfinkle, G.Horowitz, and A.Strominger, \prd 43 3140 1991.

\nref
M.Ortiz, \prd 45 2586 1992.

\nref
K.Lee, V.P.Nair, and E.J.Weinberg, \prd 45 2751 1992. [hep-th/9112008]

\nref
M.J.Duff and J.X.Lu, \npb 416 301 1994. [hep-th/9306052]

\nref 
A.Dabholkar and J.A.Harvey, \prl 63 719 1989.

\nref
A.A.Tseytlin, \plb 363 223 1995. [hep-th/9509050]

\nref 
E.Martinec, \cqg 12 941 1995. [hep-th/9412074]

\nref
G.T.Horowitz and R.Myers, {\it The value of singularities}, gr-qc/9503062.

\nref
E.Witten, \npb 195 481 1982.

\nref
F.Dowker, J.P.Gauntlett, G.W.Gibbons and G.T.Horowitz, {\it The Decay
of Magnetic Fields in Kaluza-Klein Theory}, hep-th/9507143.

\bye